# A method to determine formation position of comb teeth of Kerr micro-comb under the influence of $\chi^{(2)}$ and $\chi^{(3)}$ nonlinearity

## Hang Shen[a] Chaoying Zhao[a,b,*]

[a] College of Science, Hangzhou Dianzi University, Zhejiang, 310018 P. R. China;

[b] State Key Laboratory of Quantum Optics and Quantum Optics Devices, Shanxi University, Taiyuan, 030006 P. R. China

The Kerr microcomb has a huge potential advantage as a quantum computing platform because of its large-scale and globally coherent optical modes. The micro-comb has always faced a primary problem is how to increase the controllability of frequency domain modes. In this work, based on the pump thresholds of different side modes, we establish a set of method for determining the formation position of side mode comb teeth in a micro-ring cavity with second-order ($\chi^{(2)}$) and third-order ($\chi^{(3)}$) nonlinearity. Based on the second-order autocorrelation function $g^{(2)}(\tau)$ spectrum simulation, our quantum dynamical explanation has a good agreement with the experimental results.

## 1. Introduction

Kerr combs assisted by $\chi^{(2)}$ nonlinearity and in particular, those generated by microring resonators have been at the center of some of the most important research in fundamental physics over the last few decades [1]. When a resonator is pumped weakly, spontaneous parametric processes populate resonator modes in pairs. In this regime, microcombs can be a quantum resource for the generation of heralded single photons and energy-time entangled pairs [2], multiphoton entangled states [3] and squeezed vacuum states [4]. When pumped more strongly, the parametric gain can exceed the resonator loss and give rise to optical parametric oscillation (OPO) and bright comb formation. The modes of a microcomb can store and process a large amount of information [5]. This regime of microcombs operation can additionally yield higher noise robustness [6] and allow new algorithms for high-dimensional quantum computation [7].

Although the microcomb has been mostly studied classically, it is nonetheless fundamentally governed by the dynamics of quantized parametric processes: each resonator mode is coupled to every other mode through a three-photon interaction or four-photon interaction. The multimode coupled processes of the microcomb resemble those of the well-studied synchronously pumped

OPO [8]. If the quantum processes can be harnessed, microcombs may open a pathway toward the experimental realization of a multimode quantum resource [9] in a scalable, chip-integrated platform [10]. With their unique frequency signature, microcombs can enable photon entanglement in discrete frequency modes that allows for the generation and control of quantum states with considerably enhanced complexity. They achieve this in a scalable, manufacturable platform for generating frequency-multiplexed heralded photons and multiphoton, high-dimensional and hyper-entangled states. This quantum scalability comes from their ability to operate in the frequency picture where more frequency modes and higher dimensions can be easily added to the system without significant overhead or penalty. In contrast to continuous-variable systems, working with discrete-variable photon states is compatible with post-selection using advanced single-photon detection technologies, where losses do not intrinsically degrade the measured quantum state; loss does, however, result in the practical reduction of detection rates. Importantly, quantum photonic systems operating near 1,550 nm also tremendously benefit from, and can build on, well-established telecommunications technology, allowing the realization of chip-/fibre-based quantum devices and distributed fiber quantum networks [11]. A recent experimental article on the measurement of second-order coherence function $g^{(2)}(\tau)$ in Kerr resonator shows that when the pump is below the threshold, the formation position of the main comb can be determined by the $g^{(2)}(\tau)$ [12]. Therefore, we rationally infer that FPOCT can also be determined in the same way, which will be a method to realize the control of comb teeth theoretically.

The early research on FPOCT was based on the calculation of the size of the side modes stable region which is on a pump-centered spectrum, under the control of pump detuning [13]. This is called modulation instability (MI) gain. In the analysis of MI gain under the combined effect of $\chi^{(2)}$ and $\chi^{(3)}$ nonlinearity, compared with the case with only $\chi^{(3)}$ nonlinearity, the gain region will be in the middle of the spectrum or on both sides of the central frequency [14], which means that the spectrum will be broadened under the joint regulation of $\chi^{(2)}$ nonlinearity, and the corresponding phase matching region is larger. The analysis of modulation instability gain did not spy out the specific FPOCT. in [15] the dynamic equation of side modes quantum noise is obtained below the pump threshold, which only considers $\chi^{(3)}$ nonlinearity, and the side modes noise spectrum is obtained through the commutation relation. The single-peaked and double-peaked side modes noise spectra can be obtained by adjusting the pump detuning, which corresponds to different modulation instability gain sections. In the experimental research [16], the motion equation of the corresponding mode in the steady state is solved, and two hybridized dressed states spectra driven by two pump fields through strong $\chi^{(2)}$ nonlinearity are obtained.

However, all the above-mentioned explorations on FPOCT have not obtained a comprehensive formula for calculation on it. For modulation instability, FPOCT can only be judged within a possible section. The methods in [15,16] only account for only the $\chi^{(2)}$ or the $\chi^{(3)}$ nonlinearity, which have different effects on FPOCT. the Zeno effect produced by the change of the intensity ratio of the $\chi^{(2)}$ and $\chi^{(3)}$ nonlinearity is also important for FPOCT [17]. It is important to develop theoretical models for understanding the nature of FPOCT in various resonators.

In this paper, we report a systemic way to analyze FPOCT with a full map in the frequency domain to realize the generation and control of quantum state in optical resonators with both second- and third-order nonlinearities. The method is based on that the pump is at the threshold of the resonant cavity, and the threshold power of the pump with different side modes is different, through the influence of many parameters on it, the characteristics of comb tooth formation are analyzed. Furthermore, in order to get the specific position of comb teeth formation, we get the position with the strongest self-correlation through $g^{(2)}(\tau)$ of different side modes, and then get the specific calculation formula of FPOCT. In order to verify the correctness of our conclusion, we compare the simulation and experimental results of frequency spectrum with our theory, which is highly consistent. Our investigation results establish a general method to determine FPOCT in different nonlinear microring cavities, and it is suitable for the case that the pump is at the threshold, and the specific pump threshold size of comb teeth formation can be obtained. With the realization of the microcomb as a quantum computing platform step by step, our work can provide theoretical support for the multimode scalability of the modes.

## 1. Construction of Hamiltonian

We consider a microring optical resonator made of $\chi^{(2)}$ and $\chi^{(2)}$ nonlinearity materials in general, which has a high Q value. For example, as shown in Fig. 1(a), the pump is coupled into the resonant cavity through the optical fiber, and the resonator can trap photons in torus-like eigenmodes for a long time in a whispering gallery mode (WGM) [15]. The pump photons in the resonant cavity undergo full light-matter interaction in the process of OPO and non-degenerate OPO and are converted into the fundamental frequency modes and the second harmonic frequency modes through combined nonlinear effect, giving rise to microcombs centered at frequencies $\omega_0$ and $\Omega_0$ $(2\omega_0)$ [Fig. 1(b)]. the two mode families can

be represented by the Hamiltonian

$$\mathcal{H} = \sum_l \hbar\omega_l \hat{a}_l^\dagger \hat{a}_l + \sum_l \hbar\Omega_l \hat{b}_l^\dagger \hat{b}_l,\tag{1}$$

Where $\hbar$ is reduced Planck constant, $a_l$ and $b_l$ represent the bosonic operators for fundamental frequency modes and second harmonic frequency modes, respectively.

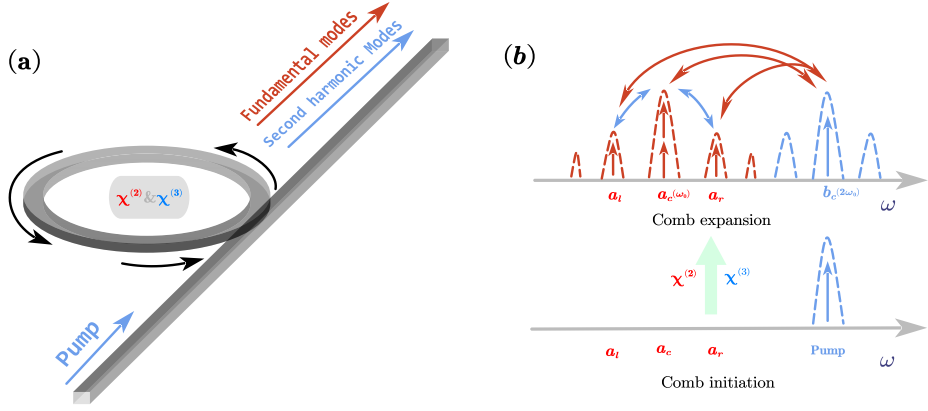

Fig.1. Schematic of the OFC generation in a ring resonator containing a nonlinear material.

We express the frequency of fundamental and second harmonic modes using Tyler expansion

$$\begin{aligned}\omega_l &= \omega_0 + d_1 l + d_2 l^2/2\\\Omega_l &= \Omega_0 + D_1 l + D_2 l^2/2\end{aligned},\tag{2}$$

$\omega_0$ ($\Omega_0$) is the angular frequencies of mode $a_c$ ($b_c$), respectively. the pump frequency is around the resonant frequency of $b_c$ and $\hbar\epsilon_{b_c}\left(\hat{b}_c^\dagger e^{-i(\omega_p + \delta_p)t} + \hat{b}_c e^{i(\omega_p + \delta_p)t}\right)$ as the pump term is added to the Hamiltonian amount, where $\epsilon_{b_c} = \sqrt{\dfrac{2k_b^{out} P_{in}}{\hbar\omega_p + \delta_p}}$ is the pump field amplitude, $\delta_p$ is detuning of the pump frequency $\omega_p$ from the adjacent resonant frequency $\Omega_0$, $P_{in}$ is pump power, $k_b^{out}$ is the external coupling coefficient. We simply assume that the coupling coefficients are uniform in the same mode family, the fundamental modes have the total coupling factor $k_a = \dfrac{\omega_0}{2Q_a}$, $Q_a$ is the quality factor about the resonator, the intrinsic loss coefficient and the external coupling coefficient $k_a^{in} = k_a^{out} = \dfrac{1}{2}k_a$. The second harmonic modes have the total coupling factor $k_b = \dfrac{\Omega_0}{2Q_b}$, and the intrinsic loss coefficient $k_b^{in} = \dfrac{1}{3}k_b$, the external coupling coefficient $k_b^{out} = \dfrac{2}{3}k_b$, the difference between the intrinsic loss coefficient and the external coupling coefficient in the total coupling factor between the two mode families is based on [19].

In order to remove the dependence of the Hamiltonian on time, through the same method in [12], we put the system into a rotating coordinate system by a unitary change matrix $\hat{U}(t) = e^{i\hat{R}t}$,

$\hat{R} = \sum_l (\omega_p + d_1 l + \delta_p) \hat{a}_l^\dagger \hat{a}_l$, which is divided by $d_1$ at equal intervals. The frequencies of fundamental and second harmonic modes are $\omega_a = \omega_0 + d_1 l + \delta_p$, $\omega_b = 2(\omega_0 + \delta_p) + d_1 l$. $d_1 = FSR \times 2\pi$ can be measured by transmission spectrum, $FSR = \dfrac{c}{2\pi n_{eff} R}$, $l$ is the number of intervals between the optical state and the frequency of the mode family center with respect to $d_1$, c is the speed of light in vacuum, $n_{eff}$ is the effective refractive index of the resonator material, and R is the radius of microring resonator. The system Hamiltonian is expressed as

$$\hat{\mathcal{H}}_{sys} = \sum_l \hbar \delta_{a_l} \hat{a}_l^\dagger \hat{a}_l + \sum_l \hbar \delta_{b_l} \hat{b}_l^\dagger \hat{b}_l + \hat{\mathcal{H}}_{\chi^{(2)}} + \hat{\mathcal{H}}_{\chi^{(3)}} + \hbar \epsilon_b \left( \hat{b}_c + \hat{b}_c^\dagger \right), \tag{3}$$

Where $\delta_{a_l} = d_2 l^2 - \delta_p$, $\delta_{b_l} = \Omega_0 + (D_1 - d_1) l + D_2 l^2 - 2(\omega_0 + \delta_p)$, $\hat{\mathcal{H}}_{\chi^{(2)}} = \sum_{lmn} \hbar g_2^{lmn} \left( \hat{a}_l^\dagger \hat{a}_m^\dagger \hat{b}_n + h.c \right)$ and $\hat{\mathcal{H}}_{\chi^{(3)}} = \sum_{klmn} \hbar g_3^{klmn} \left( \hat{a}_k^\dagger \hat{a}_l^\dagger \hat{b}_m \hat{b}_n + \hat{a}_k \hat{a}_l \hat{b}_m^\dagger \hat{b}_n^\dagger + h.c \right)$ represent the $\chi^{(2)}$ nonlinear effect and the $\chi^{(3)}$ nonlinear effect, respectively. The $\chi^{(2)}$ nonlinear effects include degenerate OPO process and non-degenerate OPO process corresponding to spontaneous parametric down-conversion (SPDC), the $\chi^{(3)}$ nonlinear effect only includes four-wave mixing (FWM), excluding cross-phase modulation and self-phase modulation.

$$\begin{aligned}
\hbar g_2 &\approx \sqrt{\frac{\hbar \omega_1 \hbar \omega_2 \hbar \omega_3}{\epsilon_0 \epsilon_1 \epsilon_2 \epsilon_3}} \chi^{(2)} \frac{1}{\sqrt{2\pi R}} \frac{1}{\sqrt{A_{eff}}} \xi_2 \\
\hbar g_3 &\approx \sqrt{\frac{(\hbar \omega_1)^2 \hbar \omega_2 \hbar \omega_3}{(\epsilon_0 \epsilon_2)^2 \epsilon_1 \epsilon_3}} \chi^{(3)} \frac{1}{2\pi R} \frac{1}{A_{eff}} \xi_3
\end{aligned} \tag{4}$$

Where $g_2$ and $g_3$ represent the second-order $\chi^{(2)}$ and third-order $\chi^{(3)}$ nonlinear coupling strengths, respectively. $A_{eff}$ is the effective mode volume, $\xi$ $(0 < \xi < 1)$ is the mode overlap factor, Equation (4) can be deduced from the Eq. (5),

$$\begin{aligned}
H^{(2)} &= \frac{1}{3} \int dV \sum_{i,j,k} \chi_{ijk}^{(2)} E_i E_j E_k = \sum_{ijk} \hbar g_2^{ijk} \hat{a}_i^\dagger \hat{a}_j^\dagger \hat{b}_k \\
H^{(3)} &= \int dV \sum_{i,j,k,l} \chi_{ijkl}^{(3)} E_i E_j E_k E_l = \sum_{ijkl} \hbar g_3^{ijkl} \hat{a}_i^\dagger \hat{a}_j^\dagger \hat{b}_k \hat{b}_l
\end{aligned}, \tag{5}$$

Through the standardized conditions of electric field $\int dV \sum \chi^{(n)} (E)^n = \hbar \omega$, the single-photon coupling strength under the influence of nonlinear effects can be quantified by Eq. (4).

## 2. Derivation of pump threshold formula with different side modes

In order to obtain the pump power at the threshold of the resonant cavity and maintain the universality of the system, we need to simplify the Hamiltonian. Under the joint influence of $\chi^{(2)}$ and $\chi^{(3)}$ nonlinear effects, the simplified Hamiltonian of the system is

$$\mathcal{H} = \hbar\delta_{a_c}\hat{a}_c^\dagger\hat{a}_c + \hbar\delta_{b_c}\hat{b}_c^\dagger\hat{b}_c + \hbar\delta_{a_l}\hat{a}_l^\dagger\hat{a}_l + \hbar\delta_{a_r}\hat{a}_r^\dagger\hat{a}_r +$$
$$\hbar g_2\left(\hat{b}_c^\dagger\hat{a}_c\hat{a}_c + \hat{b}_c\hat{a}_c^\dagger a_c^\dagger + \hat{b}_c^\dagger\hat{a}_l\hat{a}_r + \hat{b}_c\hat{a}_l^\dagger a_r^\dagger\right) + \qquad , \qquad (6)$$
$$\hbar g_3\left(\hat{a}_c\hat{a}_c\hat{a}_l^\dagger a_r^\dagger + \hat{a}_c^\dagger\hat{a}_c^\dagger\hat{a}_l\hat{a}_r\right) + \hbar\epsilon_{b_c}\left(\hat{b}_c + \hat{b}_c^\dagger\right)$$

In the fundamental modes, $a_l$ and $a_r$ are symmetrical modes centered on $a_c$, which are located on the left and right sides of $a_c$, respectively. $k_x$ is the total coupling factor for cavity mode $x$, $x \in (a,b)$. on the basis of the extended Heisenberg equations of motion

$$\dot{x} = -i\,[x, \mathcal{H}] - k_x x \qquad (7)$$

We substitute Eq. (6) into Eq. (7) to get the Heisenberg motion simplified equation of the system

$$\begin{cases} \dfrac{d}{dt}a_c = (-i\delta_{a_c} - k_a)a_c - ig_2 b_c a_c^\dagger - ig_3 a_l^\dagger a_r \\[2mm] \dfrac{d}{dt}b_c = (-i\delta_{b_c} - k_b)b_c - ig_2(a_c a_c + a_l a_r) - i\epsilon_{b_c} \\[2mm] \dfrac{d}{dt}a_l = (-i\delta_{a_r} - k_a)a_l - ig_2 b_c a_r^\dagger - ig_3 a_r^\dagger a_c a_c \\[2mm] \dfrac{d}{dt}a_r = (-i\delta_{a_c} - k_a)a_r - ig_2 b_c a_l^\dagger - ig_3 a_l^\dagger a_c a_c \end{cases} , \qquad (8)$$

We express the coherent state operator $\hat{X}$ , $X \in \{a_l, a_c, a_r, b_c\}$ representing the light field in the cavity as the sum of the average field amplitude and quantum noise: $\hat{X} = X + \delta X$. In view of the small influence of the noise term $\delta X$ at the threshold, we term in Eq. (8) and only keep the average amplitude term $X$ of the modes. When the system is in steady state, the time bias term $\frac{d}{dt}X = 0$, we get this

$$|g_2 b_c + g_3 a_c{}^2|^2 = (-i\delta_{a_c} - k_a)(i\delta_{a_c} - k_a). \qquad (9)$$

When the pump power is at the threshold of the resonant cavity, we consider that the spectral broadening in the fundamental mode family is weak because of dispersion. Therefore, under $a_l = a_r = 0$ , the intensity of $a_c$ and $b_c$ in the cavity is obtained by Eq. (8) and Eq. (9), and the pump power of the system at the threshold is obtained

$$P = \frac{\hbar(\omega_{b_c} + \delta)}{2k_b^{in}}\left(\frac{k_b{}^2(k_a{}^2 + \delta_{a}{}^2)}{g_2{}^2} + \left|\frac{g_2}{g_3}(\sqrt{(i\delta_{a_c} - k_a)(-i\delta_{a_c} - k_a)} - \sqrt{k_a{}^2 + \delta_a{}^2}) + \delta_{b_c}\sqrt{\frac{k_a{}^2 + \delta_a{}^2}{g_2{}^2}}\right|^2\right). \qquad (10)$$

Equation (10) is similar to Eq. (15) in [20], combined with $\delta_{a_l} = \delta_{a_r} = \delta_{a_c} = d_2 l^2 - \delta_p$, and $l$ is the different optical modes in the spectrum. We can get the threshold power of different $l$ in the fundamental modes by Fig. 2.

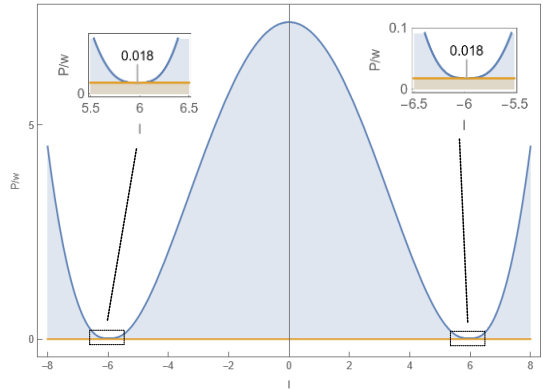

Fig.2. The threshold power of different modes in the fundamental modes

The experimental parameters are used in the following numerical simulation[19] :

$$g_2/2\pi = 0.1\,\mathrm{MHz}\,, g_3/2\pi = 1.5\,\mathrm{Hz}\,, D_1/2\pi = 3.51 \times 10^{11}\,\mathrm{Hz}\,, D_2/2\pi = 34 \times 10^6\,\mathrm{Hz}\,,$$

$$d_1/2\pi = 3.63 \times 10^{11}\,\mathrm{Hz}\,, d_2/2\pi = 20 \times 10^6\,\mathrm{Hz}\,, Q_{a_0} = Q_c = Q_d = 6.0 \times 10^5\,, Q_{b_0} = 2.0 \times 10^5.$$

We define a parameter $\delta_{OPO} = \Omega_0 - 2\omega_0 = 25k_b$. As shown in Fig. 2, $\delta_{OPO} \neq 0$ corresponds to the nondegenerate OPO process, only two low threshold positions (LTPs) appear in the side mode region. $\delta_{OPO} = 0$ corresponds to coexistence of the degenerate and the non-degenerate OPO processes[21], another LTP will appears in the middle position of Fig. 1. Our model is suitable for degenerate and non-degenerate OPO processes, the change of phase-matched sideband can be affected by adjusting $\delta_{OPO}$. In this paper, we only discuss the nondegenerate OPO process[22]. In the case of coexistence of the degenerate and the nondegenerate OPO process, due to phase-matched sideband is larger than that of the nondegenerate OPO process[23], we can predict that the broadening of comb teeth is narrower than that of the nondegenerate OPO process. As shown in Fig. 2, when l closes to $\pm 6$, the LTPs of the side mode is symmetrical with respect to the central mode, whose pump threshold is 0.018 W, the same value used in Ref.[20]. We can fully convinced that when the pump power increases, the comb teeth appears firstly at LTPs, and then the formation of the frequency comb gradually extended to the surroundings region centered on the LTPs, which corresponds to the comb tooth forming process.

First of all, $\chi^{(2)}$ nonlinear coupling coefficient has reached an exceedingly high level through the optical mode matching technology and micro-ring processing technology in optical fiber materials [24]. And we set $g_2$ changes from $2\pi \times 0.06\,\mathrm{MHz}$ to $2\pi \times 0.1\,\mathrm{MHz}$ for observing the change of pump threshold at different $l$, keep other parameters unchanged, LTPs remain unchanged. The threshold power for LTPs decreases about 1/470. The pump power is 0.018 W, as the $\chi^{(2)}$ nonlinear effect strengthens, the LTPs gradually appears on the spectrum and the intensity increases.

Our model only takes into account the third-order nonlinear effects of the fundamental mode, we change $g_3$ from $2\pi \times 1\,\mathrm{Hz}$ to $2\pi \times 5\,\mathrm{Hz}$ according to the experimental parameters [19]. As shown in Fig. 4(b1), LTPs is almost unchanged, the pump threshold at LTPs gradually increasing, and the pump threshold of the surrounding modes getting small, the ratio between the pump threshold of LTPs and the surrounding mode changes from $0.018/3.5$ to $0.494/5$, which means that the light field energy is more evenly distribute on the fundamental mode. The energy transfers from LTPs mode to the adjacent mode, which is also consistent with the physical mechanism of FWM. With the increasing of $g_3$, the ratio of $g_2$ to $g_3$ decreases, the intensity of the second harmonic mode increases due to Zeno effect [17], and the fundamental mode weakens due to energy conservation. When $g_3$ is lower than $2\pi \times 1.5$ Hz, the number of LTPs changes from 2 to 4, which already implied in Fig. 2. The $g_3$ is small enough, the two LTPs will be more concentrated in the central mode. we predict that the larger the $g_3$, the larger the phase matching bandwidth, the two LTPs far away from the central mode will gradually move out, and the same phenomenon also appears in the process of the pump detuning, which can be verified by experiment.

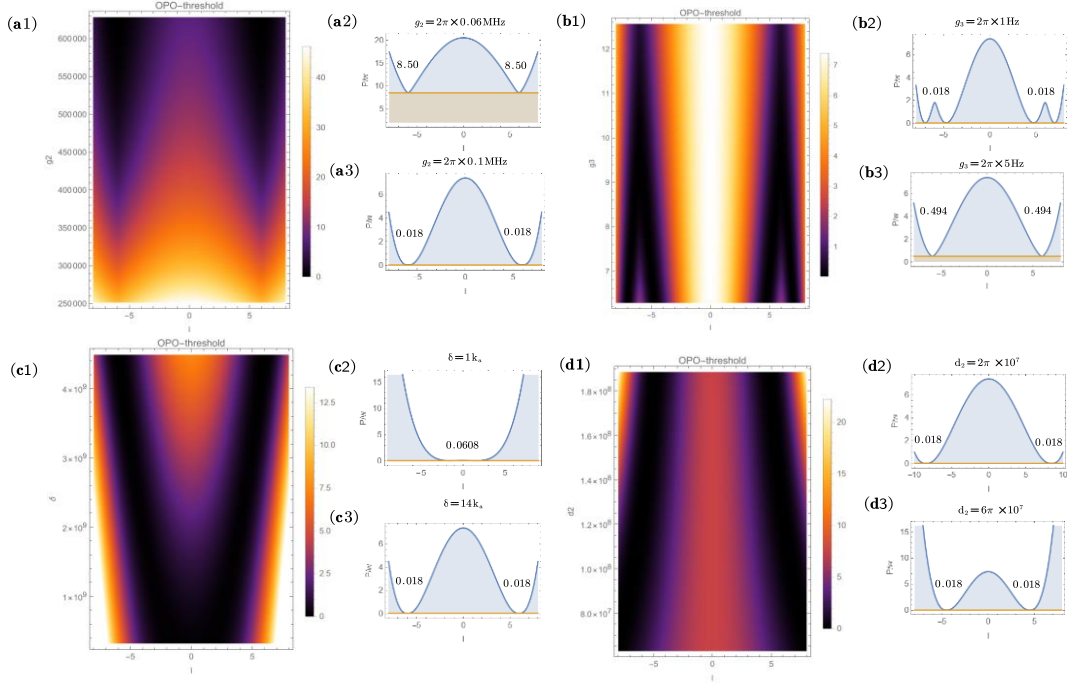

Fig.3. The changes of pump threshold at different L are obtained by adjusting g2, g3, δ and d2.

What's more, according to the simulation method in Ref.[22], we gradually increase the pump detuning from 1ka to 20ka, Fig. 3(c1) shows that LTPs gradually move away from the center mode by 0.34 FSR/ka, this speed will slow down with the detuning. It is interesting that the threshold power of the optical mode corresponding to LTPs first decreases and then increases, and when the detuning takes 14 ka, there is a minimum pump threshold of 0.018w at l=6. We find that in the infrared basic mode family[18], the OPO detuning adjustment process corresponds to 1520nm-1600nm, the symmetrical peak just occurs on both sides of 1560nm, the center mode is 6.15FSR, and the difference with our simulation results can be attributed to the fact that our model does not consider the third-order effect between the two modes and the larger power used for pumping in experiment.

In the process of detuning and amplification, the changes similar to SMLTPs in $g_3$, which corresponds to the multimodal appearance in experimental results. The intensity of outer bimodal is weak, due to the corresponding mode loss is large. There is a relationship between first-order dispersion and second-order dispersion:

$$d_2 = -\frac{c}{n_{eff}} d_1^2 \beta_2$$

The second-order dispersion expression $d_2 = -\frac{c^3}{n_{eff}^3 R^2} \beta_2$ can be derived from the first-order dispersion $d_1 = \frac{c}{n_{eff} R}$. We gradually increase $d_2$ from $2\pi \times 10^7$ to $6\pi \times 10^7$, which closers to the center frequency, the corresponding pump threshold remains at 0.018W.

The size of micro-ring cavity changes from $1.16\,\mu m$ to $1.21\,\mu m$ in experiment[19]. The larger is the size of micro-ring cavity, the lower is its $d_2$, $g_2$ and $g_3$, we can study the spectrum broadening by investigating the influence of $d_2$ on LTPs, which make LTPs far away from the center mode.

## 3. Frequency comb spectrum simulation of the fundamental modes

In the following, we want to simulate the frequency spectrum of the fundamental modes. The dynamic equation of the system can be obtained through Eq.(3) and Eq.(7).

$$\frac{d}{dt}a_l = (-i\delta_a - \kappa^a)a_l - i\sum_{k,l} 2g_2^{lmn} a_m^\dagger b_n - i\sum_{k,l,n} 2g_{3aa}^{klmn} a_k^\dagger a_m a_n - i\sum_{k,l,n} g_{3ab}^{klmn} a_k b_l^\dagger b_m,$$

$$\frac{d}{dt}b_l = (-i\delta_b - \kappa^b)b_l - i\sum_{k,l} g_2^{lmn} a_m a_n - i\sum_{k,l,n} 2g_{3bb}^{klmn} b_k^\dagger b_m b_n - i\sum_{k,l,n} g_{3ab}^{klmn} a_m^\dagger a_k b_m - i\epsilon_p \delta_D(l - b_c)$$

(11)

Where $g_{3aa}^{klmn}$ is the third-order single-photon coupling coefficient in the fundamental modes, $g_{3bb}^{klmn}$ is the third-order single-photon coupling coefficient in the second harmonic modes, $g_{3ab}^{klmn}$ is the third-order single-photon coupling coefficient between the fundamental modes and the second harmonic modes, $\delta_D(l - b_c)$ indicates that the pump is added to $b_c$ mode. We use the Four-order- Runge-Kutta method in Ref[25], and obtain the time evolution frequency comb spectrum. The pump power is 0.018w, and combined with the experimental parameters in Fig. 2 to obtain the frequency comb spectrum of the basic mode family, it is obvious that the comb teeth appear around l=6 in Fig. 5(b), As we expected, the threshold low point in Fig. 2 is the formation position of the main comb. We keep the pump power at 0.018w, adjust $g_2$ from 0 to $2\pi \times 0.06\,\mathrm{MHz}$, and there is no comb tooth formation, when $g_2$ decreases, the pump threshold of LTPs increases, The higher we adjust the pump power, the more obviously appear comb teeth at LTPs, as shown in Fig. 5(a). When the pump power is 0.018W, we find that the shape of $g_3$ has no effect on the basic mode family spectrum, considering that $g_3$ is a third-order nonlinear effect, when the pump at threshold, the intensity of the basic mode family is weak, and the gain of $g_3$ relative to other nonlinear effects can be ignored, so we adjust the pump power to 0.04W. During the increase of $g_3$ from $2\pi \times 1\,\mathrm{Hz}$ to $2\pi \times 5\,\mathrm{Hz}$, the mode intensity near LTPs decreases by more than 20 dB, which indicating that $g_3$ enhancement makes the energy of the light field more evenly distributed between different modes, but without affects LTPs.

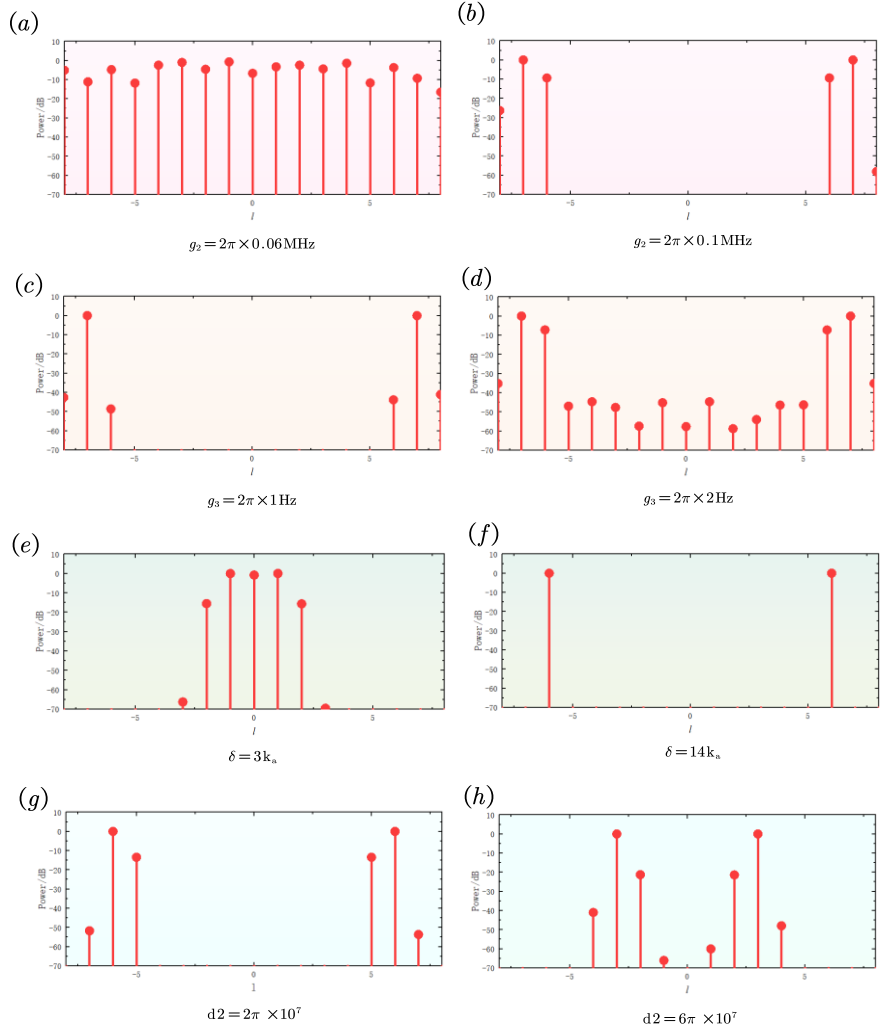

Fig. 5. Simulation of spectrum

When the pump detuning is adjusted from 1ka to 14ka, the peak position of the spectrum shift from the center to l=6, the results of spectrum simulation are consistent with that we analysis in Fig. 4(c1). When the detuning is 3ka, the corresponding frequency spectrum is similar to the degenerate OPO process in Ref.[26]. In the process of adjusting $d_2$, LTPs approaches to the center mode as $d_2$ increases, and we can see that the effect of $d_2$ and pump detuning on a is opposite, LTPs in Fig. 2 turns to FPOCT.

## 3. The calculation formula for the position of the side mold formation

We are interested in the specific formula for calculating the position of the comb tooth generation at the fundamental frequency mode when the pump is in the threshold state. We understand that the transmission spectrum of the electromagnetically induction transparency (EIT) phenomenon is similar to the image in Fig. 2, and we have shown that EIT can be achieved in the microring [27], indicating that the light field in the microring has coherent properties, this implies the existence of EIT effect in Ref. [28]. Therefore, we calculate the coherence behavior between light modes by $g^{(2)}(\tau)$. Therefore, we need to analyze the system below the threshold, when the side mode has not yet formed, we need to consider the noise effect in the mode, We express the coherent state operator representing the light field in the cavity as the sum of the average field amplitude and quantum noise: $\hat{X} = X + \delta X$, $X \in \{a_0, b_0, c, d\}$, we consider the noise term, and by the Heisenberg equation we get the quantum-coupled equation for noise, and separate the constant term from it, ignoring the influence of the higher-order term, and obtain:

$$\begin{cases} \frac{d}{dt}\delta c(t) = (-i\delta_c - k_c)\delta c(t) - ig_2|b_0|\delta d^{\dagger}(t) - ig_3|a_0|^2\delta d^{\dagger}(t) + \sqrt{2k}\,\delta c_{in}(t) \\ \frac{d}{dt}\delta d(t) = (-i\delta_d - k_d)\delta d(t) - ig_2|b_0|\delta c^{\dagger}(t) - ig_3|a_0|^2\delta c^{\dagger}(t) + \sqrt{2k}\,\delta d_{in}(t) \\ \frac{d}{dt}\delta c^{\dagger}(t) = (i\delta_c - k_c)\delta c^{\dagger}(t) + ig_2 b_0 \delta d(t) + ig_3|a_0|^2\delta d(t) + \sqrt{2k}\,\delta c_{in}^{\dagger}(t) \\ \frac{d}{dt}\delta d^{\dagger}(t) = (i\delta_d - k_d)\delta d^{\dagger}(t) + ig_2 b_0 \delta c(t) + ig_3|a_0|^2\delta c(t) + \sqrt{2k}\,\delta d_{in}^{\dagger}(t) \end{cases} \quad (12)$$

We first take the conjugate equation of the formula, and then build the system of equations in the frequency domain through the Fourier variation

$$\tilde{X}(\omega) = \frac{1}{\sqrt{2\pi}}\int_{-\infty}^{+\infty}\hat{X}(t)\,e^{i\omega t}dt\,, \quad (13)$$

and we find that in the spectral domain, Eq. (10) can be rewritten as

$$\begin{cases} -i\omega\delta c(\omega) = (-i\delta_c - k_c)\delta c(\omega) - ig_2|b_0|\delta d^{\dagger}(\omega) - ig_3|a_0|^2\delta d^{\dagger}(\omega) + \sqrt{2k}\,\delta c_{in}(\omega) \\ -i\omega\delta d(\omega) = (-i\delta_d - k_d)\delta d(\omega) - ig_2|b_0|\delta c^{\dagger}(\omega) - ig_3|a_0|^2\delta c^{\dagger}(\omega) + \sqrt{2k}\,\delta d_{in}(\omega) \\ -i\omega\delta c^{\dagger}(\omega) = (i\delta_c - k_c)\delta c^{\dagger}(\omega) + ig_2|b_0|\delta d(\omega) + ig_3|a_0|^2\delta d(\omega) + \sqrt{2k}\,\delta c_{in}^{\dagger}(\omega) \\ -i\omega\delta d^{\dagger}(\omega) = (i\delta_d - k_d)\delta d^{\dagger}(\omega) + ig_2|b_0|\delta c(\omega) + ig_3|a_0|^2\delta c(\omega) + \sqrt{2k}\,\delta d_{in}^{\dagger}(\omega) \end{cases} \quad (14)$$

In combination with the output-output relationship

$$\hat{X}_{out} = \sqrt{2\kappa_x}\,\hat{X} - \hat{X}_{in}\,, \quad (15)$$

it is easy to find that the output annihilation and creation operators obey

$$\tilde{X}_{out}(\omega) = N_{ik}(\omega)\tilde{X}_{in}(\omega)\,, \quad (16)$$

In which $\tilde{X}_{in}(\omega)$ and $\tilde{X}_{out}(\omega)$ respectively represent a $4 \times 1$ matrix, and the matrix elements respectively represent the annihilation operators of the input and output of cd mode and its Hermite conjugate term.

$$\bar{X}_{\text{out}}(\omega) = \begin{bmatrix} \hat{X}_{\text{out},c}(\omega) \\ \hat{X}_{\text{out},d}(\omega) \\ \hat{X}_{\text{out},c}^{\dagger}(-\omega) \\ \hat{X}_{\text{out},d}^{\dagger}(-\omega) \end{bmatrix} \quad \bar{X}_{\text{in}}(\omega) = \begin{bmatrix} \hat{X}_{\text{in},c}(\omega) \\ \hat{X}_{\text{in},d}(\omega) \\ \hat{X}_{\text{in},c}^{\dagger}(-\omega) \\ \hat{X}_{\text{in},d}^{\dagger}(-\omega) \end{bmatrix} \quad , \tag{17}$$

$N(\omega)$ represents a $4 \times 4$ matrix, which links the input annihilation operator with the output annihilation operator, thus realizing the establishment of the corresponding relationship

$$N_{ik}(\omega) = \begin{pmatrix} \dfrac{-(\delta+i\kappa)^2+\omega^2+\varsigma^2}{\delta^2+(\kappa-i\omega)^2-\varsigma^2} & 0 & 0 & -\dfrac{2i\kappa\varsigma}{\delta^2+(\kappa-i\omega)^2-\varsigma^2} \\ 0 & \dfrac{-(\delta+i\kappa)^2+\omega^2+\varsigma^2}{\delta^2+(\kappa-i\omega)^2-\varsigma^2} & -\dfrac{2i\kappa\varsigma}{\delta^2+(\kappa-i\omega)^2-\varsigma^2} & 0 \\ 0 & \dfrac{2i\kappa\varsigma}{\delta^2+(\kappa-i\omega)^2-\varsigma^2} & \dfrac{-(\delta-i\kappa)^2+\omega^2+\varsigma^2}{\delta^2+(\kappa-i\omega)^2-\varsigma^2} & 0 \\ \dfrac{2i\kappa\varsigma}{\delta^2+(\kappa-i\omega)^2-\varsigma^2} & 0 & 0 & \dfrac{-(\delta-i\kappa)^2+\omega^2+\varsigma^2}{\delta^2+(\kappa-i\omega)^2-\varsigma^2} \end{pmatrix}, \tag{18}$$

Where $\delta = \delta_c = \delta_d$, $\varsigma = g_2|b_0| + g_3|a_0|^2$.

Based on the formula of photon second-order coherence function [29]

$$g_{ij}^{(2)}(t+\tau,t) = \frac{\langle \hat{X}_{\text{out},i}^{\dagger}(t)\hat{X}_{\text{out},j}^{\dagger}(t+\tau)\hat{X}_{\text{out},j}(t+\tau)\hat{X}_{\text{out},i}(t)\rangle}{\langle \hat{X}_{\text{out},i}^{\dagger}(t)\hat{X}_{\text{out},i}(t)\rangle\langle \hat{X}_{\text{out},j}^{\dagger}(t+\tau)\hat{X}_{\text{out},j}(t+\tau)\rangle}, \tag{19}$$

the frequency domain and the time domain are related to the annihilation operator of the output through the inverse Fourier transform, and the input and output annihilation operators are related through Eq. (14)

$$\hat{X}_{\text{out},i}(t) = \frac{1}{\sqrt{2\pi}}\int_{-\infty}^{\infty} d\omega\, e^{-i\omega t}\hat{X}_{\text{out},i}(\omega) = \frac{1}{\sqrt{2\pi}}\int_{-\infty}^{\infty} d\omega\, e^{-i\omega t}\sum_{k=1}^{4}\left(N_{ik}(\omega)\hat{X}_{\text{in},k}(\omega) + N_{i(k+2)}(\omega)\hat{X}_{\text{in},k}^{\dagger}(-\omega)\right), \tag{20}$$

The vacuum fluctuations, which are necessary to avoid a violation of the Heisenberg uncertainty principle, have the following correlation properties

$$\langle \hat{X}_{\text{in},i}(\omega)\hat{X}_{\text{in},j}^{\dagger}(-\omega')\rangle = \delta_{ij}\delta(\omega+\omega') \quad \langle \hat{X}_{\text{in},i}^{\dagger}(-\omega)\hat{X}_{\text{in},j}(\omega')\rangle = 0 \tag{21}$$

A concrete form of the second-order coherence function is obtained by substituting Eq. (18) and Eq. (19) into Eq. (20)

$$g_{cc}^{(2)}(\tau) = g_{dd}^{(2)}(\tau) = 1 + \frac{e^{-2k\tau}}{\lambda^2}\left[\kappa\sinh(\lambda\tau) + \lambda\cosh(\lambda\tau)\right]^2, \tag{22}$$

where $\varsigma = g_2|b_0| + g_3|a_0|^2$, $\lambda = \sqrt{\varsigma^2 - \delta^2}$.

Eq.(20) is consistent with the formula of Ref. [7]. Ref.[7] shows that the formation position of the maximum coherence main comb. Our model shows the position of the maximum self-coherence comb teeth appear.

$$\frac{(g_2|b_0| + g_3|a_0|^2)^2 - (d_2l^2 - \delta)^2}{k^2} \to 1^- \tag{23}$$

Compared with the conclusions in Ref. [] and [], our results take into account both $\chi^{(2)}$ and $\chi^{(2)}$

nonlinearity, losses, which consistent with the experimental results.

Taking the experimental parameters, we can obtain the frequency spectrum after 5000 revolutions of light field.

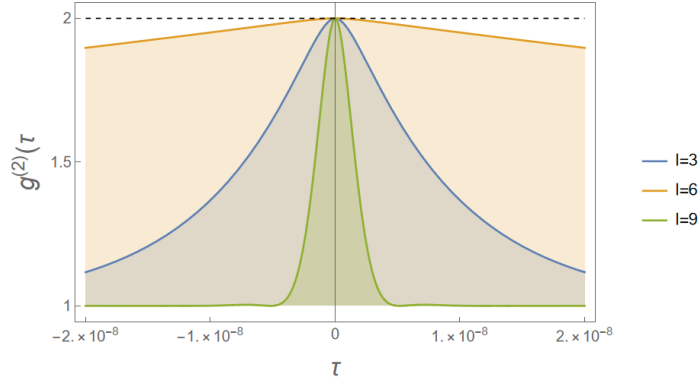

Fig. 6. The second-order coherence functions of different modes.

Bring a0 and b0 into Eq.(21), we can calculate l=6. The side peaks appear in the harmonic frequency mode family and there are 6FSR between 1600nm and 1560nm can be reasonably explained by our model. Because the pump is located in the harmonic frequency mode family, we only consider the third-order nonlinear effect. We continuous to modify Eq. (22) ,

$$\frac{(g_3|a_0|^2)^2 - (d_2l^2 - \delta)^2}{k^2} \to 1^-  \tag{24}$$

Here, the wavelength of the pump is 779.5nm, we have l=9.58. We find that there are 9.28FSR between the 780nm and 795nm.

**Conclusion**

Through the simplified model, we study the pump threshold of different side modes under the influence of $\chi^{(2)}$ and $\chi^{(2)}$ nonlinearity, and find that there will be two symmetrical low threshold positions, and through the adjustment of OPO detuning, the degenerate OPO process and the non-degenerate OPO process can be realized. For further study, we explore the influence of different physical quantities on LTPs and verify our theoretical analysis by simulating the spectrum. Inspired by the EIT effect, we explored the coherence of optical modes to determine the deterministic expression for the low threshold position, which considered more physical quantities and was more accurate in Equation 1 than in previous research, and explained the remaining problems in the article. During the validation process, it was also revealed that our model can be applied above the threshold, and our study provides a comprehensive discussion of the location of the comb teeth, which provides a possible theoretical basis for the manipulation of the comb teeth, but our model does not take into account factors such as SPM, XPM and Raman effects, and can be further refined in subsequent work.

## Referernce